\documentclass[preprint]{aastex}

\received{} 
\accepted{} 
\journalid{}{} 
\articleid{}{} 
\shortauthors{Kasen, D.}
\shorttitle{$Analytic Inversion$}
\newcommand{\rph}{\ensuremath{r_\mathrm{ph}}}
\begin{document}

\title{A Complete Analytic Inversion of Supernova Lines in the
Sobolev Approximation}

\author{Daniel Kasen\altaffilmark{1,2}\email{dnkasen@panisse.lbl.gov} 
David Branch\altaffilmark{2,1}\email{branch@mail.nhn.ou.edu} 
E. Baron\altaffilmark{2,1}\email{baron@mail.nhn.ou.edu} 
and\\
David Jeffery\altaffilmark{3}\email{jeffery@kestrel.nmt.edu}}

\altaffiltext{1}{Lawrence Berkeley National Laboratory, Berkeley, CA 94720}

\altaffiltext{2}{Department of Physics and Astronomy, University of Oklahoma, 
Norman, OK 73019}

\altaffiltext{3}{Department of Physics, New Mexico Tech, Socorro, NM 87801}
%\affil{}

\begin{abstract} 
We show that the shape of P-Cygni line profiles of photospheric 
phase supernova can be analytically inverted to extract both 
the optical depth and source function of the line -- i.e. all
the physical content of the model for the case when the Sobolev
approximation is valid.  Under various simplifying assumptions, 
we derive formulae that give $S(r)$ and $\tau(r)$
in terms of derivatives of the line flux with respect to wavelength.  
The transition region between the minimum and maximum of the line
profile turns out to give especially interesting information 
on the optical depth near the photosphere.
The formulae give insights into the relationship between line shape
and physical quantities that may be useful in interpreting
observed spectra and detailed numerical calculations. 

\end{abstract}


\section{Introduction}

The Sobolev approximation \citep{sob60,castor70,rybhum78} allows for a
simplified solution to the radiative transfer equation in media with
high velocity gradients, such as supernovae.  The Sobolev
approximation has been used to calculate synthetic spectra in stars
with strong winds \citep{CN72,ppk86} and to fit observed supernova
spectra and place constraints on explosion models
\citep{bran81b,jb90,mazz92,deng00}.  Typically these models assume
spherical symmetry and ignore continuous opacity.  Despite the
simplifying assumptions, the synthetic spectra fit the data quite
well.

To calculate line profiles in the Sobolev case two physical quantities must
be specified: (1) $\tau(r)$: the optical depth of a line
as a function of radius (often assumed to be a power law), and
(2) $S(r)$: the source function of the line as a function of radius 
(often assumed to be a resonant scattering source function).  
The Sobolev approximation has most often been used, like more sophisticated
radiative transfer techniques, 
in a direct analysis of data, where the physical quantities are
specified or calculated and the resulting spectrum is compared 
to observed data.
On the other hand, there has been some interest in taking the inverse
approach, i.e. using the line shape of observed data to infer the physical
conditions in the atmosphere.  \citet{fc89} showed that the emissivity could be
calculated for forbidden lines by differentiating the observed line
profile with respect to wavelength and estimated the effects of
electron scattering. \citet{IH00}, using the Sobolev approximation,
derived an analytic formula that gave a combination of $S(r)$ and
$\tau(r)$ as a function of the derivative of the red side of an
emission feature of arbitrary optical depth.  The run of the optical
depth of a line is then given if one specifies a form for the source
function.  For instance $S(r)$ in the case of pure resonance
scattering is given by:
\begin{equation}
S(r) = I_\mathrm{ph} W(r), \label{resscat}
\end{equation}
where
\begin{equation}
W(r) = \frac{1}{2}\biggr(1-\sqrt{1 -
\biggl(\frac{\rph}{r}\biggr)^2}\biggl)
\label{dilfactor}
\end{equation}
is the dilution factor \citep{mihalas78sa}.  $I_\mathrm{ph}$ is  
the intensity from the photosphere and $\rph$ the radius of the photosphere.

However, in supernova atmospheres the source function may deviate
strongly from pure resonance scattering.  In fact the failure
of Sobolev models to properly 
fit the shape of some spectral lines -- in particular net emission
features -- is basically because the source function is usually
assumed that of pure resonant scattering, not because of
direct limitations in the 
Sobolev approximation itself.  The source function is an
interesting quantity in its own right and ideally an inversion would
extract both $S(r)$ and $\tau(r)$ from the line profile.  In what
follows we show how the degeneracy found by \citet{IH00} can be broken
for photospheric phase supernovae and derive analytic 
expressions for both $\tau(r)$ and $S(r)$ using the shape of the
entire line profile, thus providing a complete solution to the
inverse problem.

We derive the formulae assuming spherical symmetry, an expanding line-scattering
atmosphere surrounding a sharp continuum-emitting photosphere that
absorbs any flux scattered back onto it, no continuous opacity, and no
line blending.  Even when these assumptions are not strictly valid the
formulae should still give considerable insight into the physical
conditions in the atmosphere.  On the
other hand, the limitations of the formulae provide an interesting
result in their own right -- they clearly show what type of features
are impossible under the above assumptions, making it is obvious where
more complicated scenarios must be invoked to explain a spectrum.

\section{The Sobolev Approximation\label{sobsec}}

In the most often used Sobolev model, one begins with a perfectly
sharp, spherical photosphere that emits radiation as a blackbody, 
and which is surrounded
by a moving atmosphere with large velocity gradient.
The basic idea behind the Sobolev approximation is that a photon
emitted from the photosphere 
only interacts with a line in the small region of the
atmosphere where the photon is Doppler shifted into resonance with
the line.  Since the source function can be assumed to be constant
over this small resonance region, the solution of the radiative transfer
equation is greatly simplified.  It also becomes simpler to visualize
line formation, as the line flux at a given wavelength comes 
from a 2-dimensional resonance surface in the atmosphere.
The criterion for the validity of the Sobolev approximation 
is that the resonance regions be sufficiently small,
and this is characterized
by the ratio of the atmosphere's thermal velocity 
(or mean microturbulent velocity, if significant) to the 
velocity scale height (i.e the velocity range over which temperature,
density, and occupation numbers change by a factor of order 2
\citep{jeff93}).   Quantitative accuracy is found for velocity ratios
$ \lesssim 0.1$ \citep{olson82}.  For photospheric phase supernovae 
thermal velocities are of order 10 km $\mathrm{s^{-1}}$ and
velocity scale heights are $\sim 10^3$ km $\mathrm{s^{-1}}$, giving a ratio 
of $\sim 10^{-2}$.  

\begin{figure}
\begin{center}
\leavevmode
\plotone{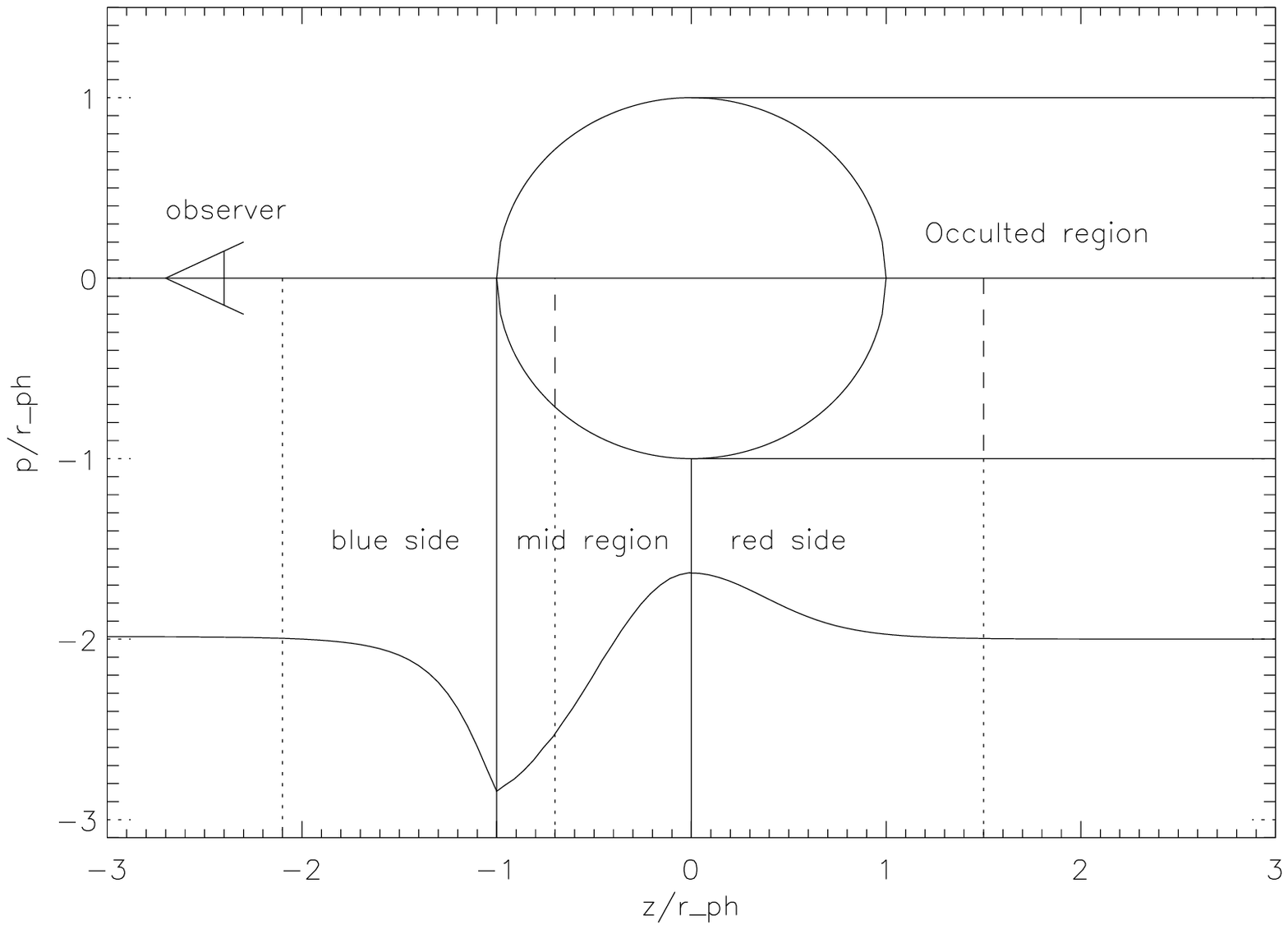}
\caption{Schematic diagram of how line profiles are
calculated in the Sobolev approximation.  The figure is a
cross-sectional view of the supernova with the sphere in the center
representing the photosphere.  The dotted/dashed lines show the region
of integration for three points on the line profile, one each on the red
side, the mid region, and the blue side.  In each case the integration
over that region of the atmosphere produces the flux at the wavelength
where the dotted line intersects the line profile.  On the mid and red
side, the dashed lines represent the region of the atmosphere where
light comes directly from the photosphere.\label{fig1}}
\end{center}
\end{figure}

For supernova atmospheres, homology is established 
shortly after the explosion, so that $v = r/t$, where $t$ is
the time since explosion.   
In this case the resonance surface for a wavelength is a
plane perpendicular to the line of sight.  We label the line
of sight with a coordinate z, with origin at the center of
expansion and with $z$ increasing away from the 
observer (see Figure~\ref{fig1}).  
A plane at coordinate $z$ has a z-component velocity 
of $v_z = z/t = z\,v_\mathrm{ph}/\rph$
where $v_\mathrm{ph}$ is the velocity of the photosphere.  This plane
is then responsible for the line flux at a wavelength 
$\lambda = \lambda_0[1 + (z/\rph\,)(v_\mathrm{ph}/c)]$, i.e  the
rest wavelength, $\lambda_0$, Doppler shifted by the z-component velocity.

The flux at a given wavelength is calculated by integrating
over all the characteristic rays of the corresponding plane.
Figure~\ref{fig1} shows that the formation of the line profile breaks 
up schematically into three regions.  
For $z \ge 0$, the flux is redshifted with respect to the line center so
we call this the red side. This leads to an expression for the flux
as an integral over impact parameter p (the coordinate perpendicular
to the line of sight):
\begin{eqnarray}
\frac{F(z)}{2\pi} & =&
       \int_{0}^{\rph}I_\mathrm{ph}p\,dp +
       \int_{\rph}^{\infty}S(r)(1-\zeta(r))p\,dp \label{redflux} \\\nonumber 
 && = \frac{1}{2} \rph^2I_\mathrm{ph} + \int_{\rph}^{\infty}S(r)(1-\zeta(r))p\,dp
\end{eqnarray}
where $\zeta(r) = e^{-\tau(r)}$
and $F(z)$ is the observed flux (apart from a factor of $1/D^2$, where
$D$ is the distance to the supernova) at wavelength 
$\Delta\lambda = \lambda - \lambda_0 = \lambda_0 zv_\mathrm{ph}/c\rph = \lambda_0z/ct$.
The first term in Eqn.~(\ref{redflux})
accounts for the flux coming directly from the photosphere, 
and the second term for photons scattered or created to emerge along the line of site.
We have assumed for 
convenience an infinite atmosphere although none of our results is
altered in the case that the atmosphere terminates at some radius
$r_\mathrm{max}$. 

For $z<0$, the integral has three terms in general, with the third
term in Eqn.~(\ref{bluesob}) now representing the region where material 
intervening between the photosphere and the observer
leads to absorption of the continuum radiation:
\begin{eqnarray} \label{bluesob}
\frac{F(z)}{2\pi} &=& \int_{0}^{p_0}I_\mathrm{ph}p\,dp
     +  \int_{p_0}^{\infty}S(r)(1-\zeta(r))p\,dp
     +  \int_{p_0}^{\rph}I_\mathrm{ph}\zeta(r)p\,dp
       \label{blueflux} \\\nonumber
 &=& \frac{1}{2}p_0^2 I_\mathrm{ph}
     + \int_{p_0}^{\infty}S(r)(1-\zeta(r))p\,dp
     +  \int_{p_0}^{\rph}I_\mathrm{ph}\zeta(r)p\,dp.
\end{eqnarray}
The limit $p_0$ is given by the $p$ location of the 
spherical photosphere for a given $z$, namely 
\[
p_0 = \left\{ \begin{array}{ll}
\sqrt{\rph^2 - z^2} &\mathrm{for}\ -\rph < z < 0\\
     0 & \mathrm{for}\ z \le -\rph 
	      \end{array}
\right.
\]
and the first term of Eqn.~(\ref{blueflux}) is identically zero for
$z \le -\rph$. We call the part of the line profile where $z < - \rph$
the blue side and the part where $- \rph<z<0$ the mid region.

\section{The Inversion Formulae}
\subsection{Inversion for $\zeta(r)$ for $\rph<r<\sqrt{2}\rph$}

We consider the inversion of each region of the line in turn, beginning
with the mid region.  The
mid region of the line profile turns out to be only sensitive to
the optical depth of the line near the photosphere.  Using
Eqn.~(\ref{blueflux}), we 
change the integration variable from $p$ to $r = \sqrt{p^2 + z^2}$,
and divide through by $I_\mathrm{ph}$:
\begin{eqnarray}
\frac{\rph^2}{2}f(z) &=& \int_{|z|}^{\rph}r\,dr
     +  \int_{\rph}^{\infty}s(r)(1-\zeta(r))r\,dr
     +  \int_{\rph}^{\sqrt{z^2+\rph^2}}\zeta(r)r\,dr
       \label{midr}, \\\nonumber
\end{eqnarray}
where we have defined
$s(r) = S(r)/I_\mathrm{ph}$ and $f(z) = F(z)/(\pi I_\mathrm{ph} \rph^2)$ (i.e. the total flux divided by the continuum flux).  
$I_\mathrm{ph}$ has been assumed to be constant over the line profile.

Written this way we see that the 
term involving the source function is independent of $z$ and so contributes a
constant amount to the flux for every point in the mid region.  
The derivative of the mid region is therefore
independent of the source function.
The change in flux from a velocity surface at $z$ to one at $z - \Delta z$ 
is due only to
the fact that a bit more of the photosphere is now obscured by
the optical depth of the line.  
One then expects the derivative $\frac{df}{dz}$ to
depend only on the optical depth.  

Since the terms in Eqn.~(\ref{midr}) only depend on $z$ in
the limits of the integral we can differentiate the integrals using
Leibnitz' rule:
\begin{equation}
\frac{d}{dz}\int_{\xi(z)}^{\eta(z)}g(t)\,dt 
= g(\eta)\frac{d\eta}{dz}  - g(\xi)\frac{d\xi}{dz}  \label{liebniz}
\end{equation}
Applying Eqn.~(\ref{liebniz}) to Eqn.~(\ref{midr}) allows us to solve for
$\zeta(r)$: 
\begin{equation}
\zeta(r=\sqrt{\rph^2 + z^2}) 
                             = 1 - \frac{\rph^2}{2 |z|}\frac{df}{dz} \\
= 1 - \frac{\lambda_0^2}{2 |\Delta\lambda|}\frac{df}{d\Delta\lambda}
\biggr(\frac{v_\mathrm{ph}}{c}\biggl)^2 \label{midbeta}
\end{equation}
which is valid for $-\rph<z<0$.  In using Eqn.~(\ref{midbeta})
 to calculate $\zeta(r)$ from
a spectrum, one can choose either $\Delta\lambda$,
$z$, or $r$ as the independent parameter.  For instance, from 
$\Delta\lambda$ (which is always less than zero for Eqn.~[\ref{midbeta}])
the other two parameters are determined by 
$z=\rph (\Delta\lambda/\lambda_0)(c/v_\mathrm{ph})$ and 
$r = \sqrt{\rph^2 + z^2}$.  The photospheric radius is itself given
by $\rph = v_\mathrm{ph} t$; however if the time since explosion is not known,
one can still determine $\zeta$ as a function of the scaled distance
$r/\rph$.

Eqn.~(\ref{midbeta}) gives us some immediate insight into the
relationship between 
line shape and optical depth.  The steepness of the mid
region (once the photospheric velocity has been scaled out)
is a direct indication of the Sobolev optical depth.
If no line feature exists, then
$\frac{df}{dz} = 0$ and hence $\zeta = 1$ (i.e $\tau = 0$).  
Thus the absence of a
feature implies either negligible line optical depth or the breakdown
of our assumptions -- in this formalism there is no choice for
the source function that allows a line to
``erase'' itself.  
A stair-step mid region could be a signal
that the optical depth near the photosphere is oscillating between
small and large values (i.e. the medium is clumpy).

Note that since $\zeta \le 1$, Eqn.~(\ref{midbeta}) implies that the derivative
$\frac{df}{dz}$ is always greater than or equal to zero; i.e. the
mid region always increases (or is flat) to the red.  
The appearance of a rising hump in the mid region, 
could indicate that the stated assumptions do not hold.

The fact that $\frac{df}{dz} \ge 0$ also implies that an emission
feature cannot peak blueward of its rest wavelength (at the most it
can remain flat into the mid region as one would have for a
detached atmosphere).  However, the peaks of emission features are
indeed found to be blue-shifted, both in real data and in spherically
symmetric NLTE models.
\citet{jb90} and \citet{duschetal95} attribute the blueshift
to an NLTE effect where a large source function near the photosphere
enhances the flux in the mid region.
Under our assumptions this cannot be correct, since Eqn.~(\ref{midr})
shows that the shape of the mid region is independent of the
source function. 
Various second order physical effects, not included in the
inversion formulae, could possibly explain the blueshifts, for example: 
continuous opacity added to the model
could preferentially extinguish photons from the red side of the envelope; 
an absorption from another line to the red could cut into the emission
peak;
a large slope in the continuum could shift the peak;
clumpiness of the photosphere \citep{wanghu94} could break the
spherical symmetry of the problem; relativistic effects can cause
a significant blueshift for high photospheric velocities 
($\ge$ approximately 15,000 km/s \citep{jeff93}); or 
line-scattered light could be diffusely 
reflected off and blueshifted by the photosphere, causing
a blueshift of the emission peak \citep{Chugai87A88}.

A few points must be made concerning the applicability of 
Eqn.(\ref{midbeta}): (1) near the rest wavelength 
the equation may not yield reliable results, since the $\Delta\lambda$
in the denominator goes to zero and must be delicately canceled by
the flux derivative also going to zero.  
Thus any noise in the flux derivative 
(which must be evaluated numerically from the data) 
will be inflated at small $\Delta\lambda$.
(2) one can not extract realistic values for $\tau \gg
1$ since $\zeta$ depends exponentially on $\tau$.  One will know
$\tau$ is large but not its exact value;  (3) at late times
(the nebular phase) the photosphere may become negligibly small and so
there is no mid region -- in this case, as we will see, our analysis 
reduces to that of
\citet{IH00};   (4) Eqn.~(\ref{midbeta}) only gives the value of $\zeta$
for the radial region 
$\rph<r<\sqrt{2}\rph$.  This is expected to be the region of highest
density opacity in the atmosphere and so our formula for $\zeta(r)$ is interesting
in itself.  For example, if an atmosphere's density scales like $r^{-8}$ 
(a density law often used in spectral analysis; e.g. \cite{millard94i99}) 
then at $r = \sqrt{2}\rph$ the optical depth has already fallen to 
1/16 of its
photospheric value.  Nevertheless, in the following we
show how it is possible to extend the solution for $\zeta(r)$ to
arbitrary $r$ by using information from the blue and red sides of the
line profile.

\subsection{Inversion for $S(r)$}
We next consider the inversion of the red side of the line, which will
allow us to solve for the source function.  Changing variables in
Eqn.~(\ref{redflux}), we see that the flux is now given by a
source term plus an unobstructed photosphere term:
\begin{eqnarray} \label{redside}
\frac{\rph^2}{2}f(z) 
 &=& \frac{1}{2}\rph^2  + \int_{\sqrt{\rph^2 + z^2}}^{\infty}s(r)(1-\zeta(r))r\,dr. 
\end{eqnarray}
The second term in Eqn.~(\ref{redflux}) is a constant with respect to
$z$ since the 
photosphere is always completely unobscured for $z > 0$.  The same
technique of differentiating the integral
allows us to solve for $s(r)$:
\begin{eqnarray}
s(r = \sqrt{\rph^2 + z^2}) &=& 
-\frac{\rph^2}{1-\zeta(r)}\frac{1}{2 z}\frac{df}{dz}\label{sourceeqn}
\\
&=& -\frac{1}{1-\zeta(r)}
\frac{\lambda_0^2}{2 \Delta\lambda}\frac{df}{d\Delta\lambda}
\biggr(\frac{v_\mathrm{ph}}{c}\biggl)^2
\nonumber
\end{eqnarray}
which is valid for all $z \ge 0$ and the independent parameter can
be chosen to be any of $\Delta\lambda$, $z$, or $r$.
This is essentially the same result derived by \citet{IH00}. 
Because Eqn.~(\ref{midbeta}) together with Eqn.~(\ref{extender}) (see below)
gives $\zeta$ everywhere, Eqn.~(\ref{sourceeqn}) can be used
to determine the source function at all radii above the photosphere.
Note if $\tau = 0$ then $\zeta =
1$ and Eqn.~(\ref{sourceeqn}) is undefined -- if a line has no optical
depth it is of course 
impossible to determine its source function.  For large
optical depth, $\zeta=0$, and the shape of the red side depends on the
source function only.  Since $s \ge 0$ and $\zeta \le 1$ we must have
$\frac{df}{dz} \ge 0$ on the red side -- the red side always decreases (or
stays flat) to the red.  One cannot have humps or even a redshifted
emission peak.  Redshifts can occur due to non-Sobolev radiative transfer
effects, but these are likely to be very small for supernovae \citep{hamann81}.

\subsection{Inversion for $\zeta(r)$ for $r>\sqrt{2}\rph$}   
Finally the flux from the blue side of the profile will allow us to
extend the solution of $\zeta$ to large $r$.  The flux is given by
a source term plus a fully obstructed photosphere:
\begin{eqnarray} \label{blueside}
\frac{\rph^2}{2}f(z) = \int_{|z|}^{\infty}s(r)(1-\zeta(r))r\,dr
     +  \int_{|z|}^{\sqrt{\rph^2 + z^2}}\zeta(r)r\,dr.
 \end{eqnarray}
The same differentiation technique yields:
\begin{eqnarray}
\zeta(r=\sqrt{\rph^2 + z^2}) = \zeta(|z|)
+ s(|z|)\{1-\zeta(|z|)\} - \frac{\rph^2}{2 |z|}\frac{df}{dz},
\label{sblue}
\end{eqnarray}
which is valid for $z < -\rph$.
Making use of  spherical symmetry,  Eqn.~(\ref{sourceeqn})
can be used to replace the second term in 
Eqn.~(\ref{sblue}) with 
\begin{eqnarray}
s(|z|)(1-\zeta(|z|)) = -\frac{\rph^2}{2 z_+}\frac{df(z_+)}{dz} \label{interresult}
\end{eqnarray}
where $z_+ = \sqrt{z^2 - \rph^2}$. Combining Eqns.~(\ref{sblue}) and
(\ref{interresult}) we obtain:
\begin{eqnarray}
\zeta(r=\sqrt{z^2 + \rph^2}) &=& \zeta(r=|z|)\label{extender} \\
&&\hspace{1pt} -
\frac{\lambda_0^2}{2}\biggr(\frac{v_\mathrm{ph}}{c}\biggl)^2 \biggr\{ \biggr[\frac{1}{\Delta\lambda}
\frac{df}{d\Delta\lambda}\biggl]_{\Delta\lambda=
\frac{\lambda_0}{ct}\sqrt{z^2 - \rph^2}} +
\biggr[\frac{1}{|\Delta\lambda|}
\frac{df}{d\Delta\lambda}\biggl]_{\Delta\lambda=-\frac{|z|\lambda_0}{ct}}
\biggl\}.\nonumber 
\end{eqnarray}
where $|z|>r_{\rm ph}$ is the independent
parameter for evaluating 
$\zeta(r=\sqrt{z^{2}+r_{\rm ph}})$ from $\zeta(r=|z|)$
and ${df\over dz}$.
Given $\zeta(r)$ for
$r\in[nr_{\rm ph},\sqrt{n+1}r_{\rm ph}]$, Eqn.~(\ref{extender}) allows 
us to evaluate $\zeta(r)$ for
$r\in[\sqrt{n+1}r_{\rm ph},\sqrt{n+2}r_{\rm ph}]$
where $n\geq 1$ is an integer.
Beginning with $\zeta(r)$ for 
$r\in[r_{\rm ph},\sqrt{2}r_{\rm ph}]$, given by Eqn.~(\ref{midbeta}), 
we can in fact use Eqn.~(\ref{extender}) to find
 $\zeta(r)$ for all $r$.

\section{Discussion}
For late times when the photosphere becomes negligibly small ($\rph
\longrightarrow 0$),  the mid region disappears and Eqn.~(\ref{midbeta})
becomes meaningless.  We cannot use the reduced quantities $f(z)$ and $s(r)$
in this case.  Instead the counterpart to 
Eqns.~(\ref{redside}) and~(\ref{blueside}) is:
\begin{equation}
 {F(z)\over2\pi}=\int_{|z|}^{\infty} S(r)[1-\zeta(r)]r\,dr \label{nophot}
 \end{equation}
which implies that $F(z)$ is symmetric about $z=0$:
i.e., the line profile is symmetric about the rest wavelength.
From Eqn.~(\ref{nophot}) we derive:
\begin{equation}
S(r)[1-\zeta(r)]=-{1\over2\pi}{1\over z}{dF\over dz}
                =-{1\over2\pi}\left({\lambda_{0}\over ct}\right)^{2}
	         {1\over\Delta\lambda}{dF\over d\Delta\lambda}
\end{equation}
where $r=|z|=ct|\Delta\lambda|$.  Since the profile is symmetric
one obtains the same information using either $z \le 0$ or $z \ge 0$.
Thus without a photosphere it is not
possible to separate $S$ and $\zeta$, however the product
$S\{1 - \zeta\}$ can be determined for all radii.

\begin{figure}
\begin{center}
\leavevmode
\plotone{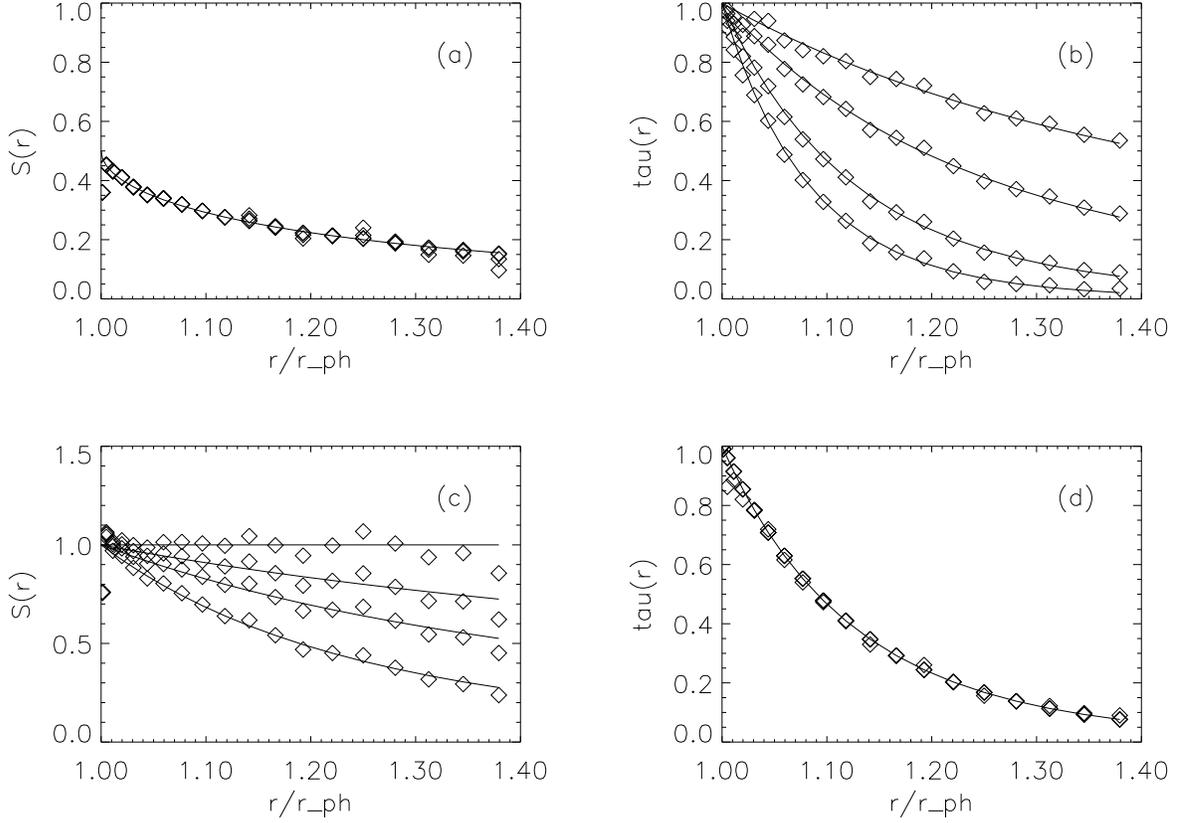}
\caption{Examples of the application of the inversion formulae.
Panels (a) and (b) show the source function and optical depth obtained
from the 
inversion of line profiles generated with the same source 
function $S = W(r)$, but different power indices ($n = 2, 4, 8, 12$) for the
optical depth. 
Panels (c) and (d) show the the source function and optical depth obtained
from the inversion of 
line profiles generated with the same optical depth power 
law ($n=8$), but different source functions ($S(r) \propto r^{-n}$,
$n=0,1,2,4$). The solid lines are the exact input functions 
and the diamonds are the functions extracted using the inversion
formulae.  The small amount of 
noise in the plots is due to numerical error.
\label{fig2}}
\end{center}
\end{figure}

To demonstrate that the above equations really do allow for a clean
inversion, we have generated line profiles 
under the given assumptions and
using $S(r)$ and $\tau(r)$ given by power laws of various exponents. 
Figure~\ref{fig2} shows that the power law behavior can be recovered by
applying the inversion formulae to the line shape.
Although here we have assumed that the functions 
are monotonically decreasing with 
$r$, this is not a necessary condition for our derivations
and the formulae apply also for
non-monotonic distributions.  

Because the inversions
in Figure~\ref{fig2} were applied to pristine model lines, the
results show very little noise (the small amount is due to
numerical error), however in application to real data, 
the quality of the inversion will of course depend upon
the signal to noise and spectral resolution of the data.
Because derivatives are especially sensitive to 
a high frequency noise component, smoothing of the spectrum
or some other stabilization technique may need to be applied, 
which typically amounts to 
assuming \emph{a priori} some level of smoothness of the 
functions $\tau(r)$ and $S(r)$ (c.f. Craig \& Brown 1986). 

\section{Conclusion}

Eqns.~(\ref{midbeta}),~(\ref{sourceeqn}), and~(\ref{extender}), taken
together constitute a complete analytic inversion of supernova
lines in the Sobolev approximation.
Given the present assumptions, some of the more interesting facts are:
(1) the steepness of the mid
region reflects the size of the optical depth; (2) the absence of
a line implies negligible optical depth; (3) a jagged mid region
signals a clumpy absorbing region near the photosphere; (4) the emission
feature may have no rising humps; (5) emission
features cannot be blueshifted or redshifted 
simply by varying the line source function or optical depth.
When applied under the right circumstances, 
the formulae may provide useful information on the physical 
conditions in the atmosphere
as well as constraints on supernova explosion models.

It is also interesting that this inversion problem 
possesses a unique solution for both $S(r)$ and $\tau(r)$.  
A persistent worry in supernova modeling is that very
different physical parameters may lead to identical 
looking synthetic spectra.  
The analytic solutions above demonstrate that, at least in principle,
each different choice of $S(r)$ and $\tau(r)$ produces a 
distinct line profile (although in practice it may be impossible to 
discern the differences from noisy data).
Although for more general models the inversion will not be unique,
the success in the present case does give some support that
a good fit to a line does
indeed imply realistic physical parameters.

\acknowledgments 
We would like to thank Peter Nugent and R. Ignace for helpful comments
and suggestions.  This work was supported
in part by NSF grants  AST-9731450, AST-9986965 and NASA
grant NAG5-3505.

\clearpage

\eject

\end{document}